\documentclass[twocolumn,showpacs]{revtex4}

\usepackage{dcolumn}
\usepackage{amsmath}
\usepackage[xdvi]{graphicx}
\usepackage{ulem}
\usepackage{color}

\newcommand{\bra}{\left\langle}
\newcommand{\ket}{\right\rangle}

\newcommand{\bi}{\begin{itemize}}
\newcommand{\ei}{\end{itemize}}
\newcommand{\be}{\begin{equation}}
\newcommand{\ee}{\end{equation}}

\begin{document}
\title{Single-molecule stochastic resonance}
\author{
K. Hayashi$^{1,2}$, S. de Lorenzo$^{3}$, M. Manosas$^4$, 
J. M. Huguet$^1$ and F. Ritort$^{1,3,*}$}

\affiliation{
1) Departament de F\'{\i}sica Fonamental, Facultat de F\'{\i}sica, 
Universitat de Barcelona, Diagonal 647, E-08028, Barcelona, Spain\\
2) Department of Applied Physics, School of Engineering, Tohoku University, 
2-1-1 Katahira, Sendai 980-8577, Miyagi, Japan. \\ 
3) CIBER de Bioingenier\'{\i}a, Biomateriales y Nanomedicina, Instituto de 
Salud Carlos III, Madrid, Spain \\
4) Laboratoire de Physique Statistique, Ecole Normal Sup\'erieure, Unit\'e 
Mixte de Recherche 8550 associ\'ee au Centre National de la Recherche 
Scientifique et aux Universit\'es Paris VI et VII, 75231 Paris, France
}
\email{fritort@gmail.com}

\date{\today}
\begin{abstract}
Stochastic resonance (SR) is a well known phenomenon in dynamical systems. 
It consists of the amplification and optimization of the response of a 
system assisted by stochastic noise. Here 
we carry out the first experimental study of SR in single DNA hairpins 
which exhibit cooperatively folding/unfolding transitions under the action of 
an applied oscillating mechanical force with optical tweezers. 
By varying the frequency of 
the force oscillation, we investigated the folding/unfolding kinetics of DNA
hairpins in a periodically driven bistable free-energy potential. We 
measured several SR quantifiers under varied conditions of the experimental 
setup such as trap stiffness and length of the molecular handles used 
for single-molecule manipulation. We find that the signal-to-noise ratio 
(SNR) of the spectral density of measured fluctuations in 
molecular extension of the DNA hairpins is a good quantifier of the 
SR. The frequency dependence of the SNR exhibits a peak at a frequency value 
given by the resonance matching condition. Finally, we carried out experiments 
in short hairpins that show how SR might be useful to enhance the detection of 
conformational molecular transitions of low SNR.


\end{abstract}
 

\maketitle 

\section{Introduction}

All nonlinear systems that exhibit stochastic noise 
are susceptible to undergo stochastic resonance (SR). 
When SR is triggered, the response of a system to an 
external forcing is amplified. 
SR has been studied in a large variety of 
systems, including climate dynamics \cite{benzi1,benzi2}, colloidal 
particles \cite{simon,schmitt,ciliberto}, biological systems 
\cite{bio,McDonAb09,reviewer2}, and quantum systems 
\cite{quantum1,quantum2}. 
With the recent advent of single-molecule techniques, it is nowadays 
possible to measure SR at the level of individual molecules. Biomolecules 
exhibit rough and complex free energy landscapes that determine folding 
kinetics 
and influence the way they fold into their native structures. The use of 
force spectroscopy techniques has become important practice in studies of 
molecular biophysics.  By applying a mechanical force at both extremities 
of an individual molecule and by recording the time evolution of the molecular 
extension (the reaction coordinate in these experiments), information 
about the folding reaction can be obtained. The application of forces makes 
possible to disrupt the weak bonds that hold their native structure  
to reach a stretched unfolded conformation.
In this way thermodynamics (e.g. the free energy of folding) and kinetics 
(the rates of unfolding and folding) can be determined.

Although most SR studies use temperature as a tunable parameter, this
is not the best choice to investigate SR effects at the
single-molecule level.  Biomolecules have a strong sensitivity to
temperature variations. Indeed, beyond increasing thermally
  assisted noise, temperature also modifies  the
shape of the molecular free energy landscape.  Thus, another tunable
parameter such as the oscillation frequency of force might be 
 more appropriate to study SR in
biomolecules. SR appears as a maximum in the response of  a biomolecule 
at a characteristic frequency (the resonance  
frequency). This occurs when a characteristic timescale of the signal (e.g. 
its decorrelation or relaxation time) matches half period of the oscillation 
(the so-called matching condition). The matching condition must not be taken 
as a strict equality but a qualitative relationship between the two 
timescales \cite{hanggi,wellens}. This means that different SR quantifiers 
may not give coincident resonance frequencies specially for low quality 
resonance peaks. It seems important to investigate which SR 
quantifier is best suited to identify SR behavior.

\begin{figure*}
\begin{center}
\includegraphics[width=16cm]{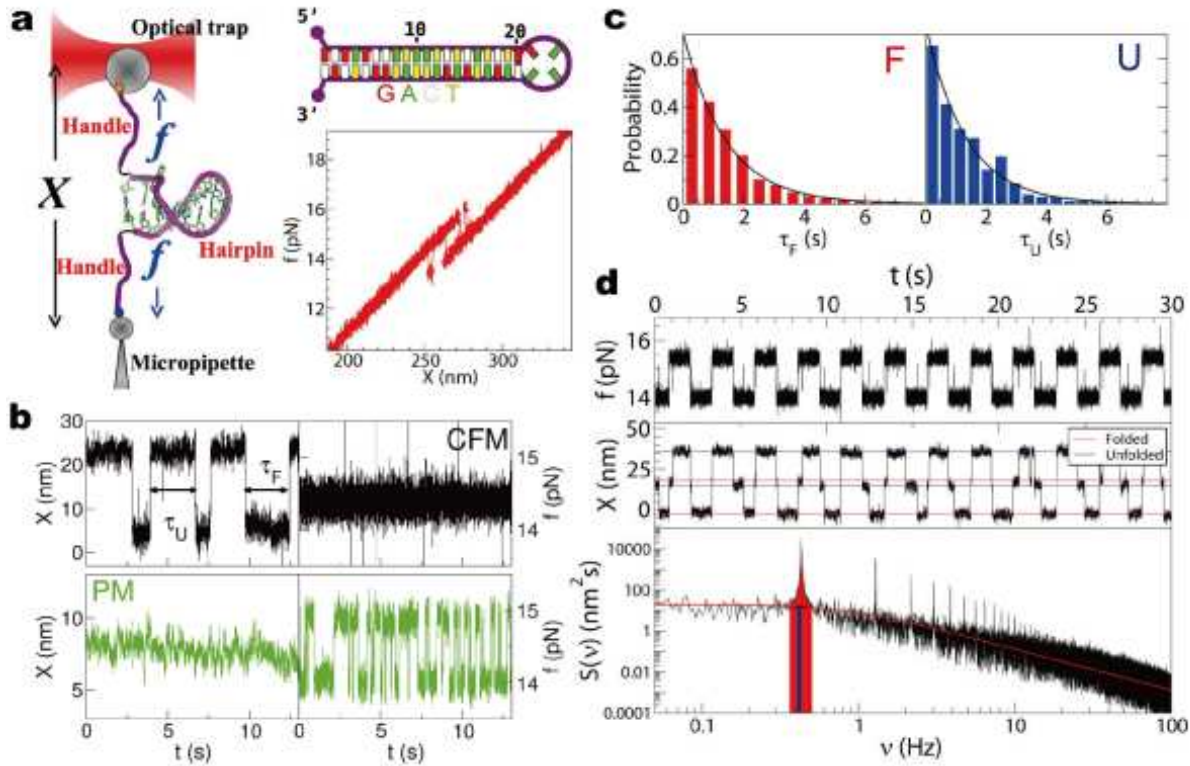}
\end{center}
\caption{ {\bf Experimental setup, hopping and SR experiments.}  ({\bf a})  
Illustration of the experimental system (left), DNA sequence of the two-state 
hairpin H1 (upper right, sequence shown in color code) and experimental 
force-distance curve for 
H1  obtained from a pulling experiment (lower right).  ({\bf b}) 
Typical force and extension traces of the hopping experiments for H1 obtained 
in CFM (upper) and PM (lower).  ({\bf c}) 
Probability distributions of 
the residence times for H1 in the F (red) and U (blue) states obtained from 
the hopping experiments at $f=f_{\rm c}\simeq 14.5$~pN in the CFM. The 
black curves show the exponential fit, $(1/a)\exp(- \tau/a)$, to the data, 
with $a$=1.42 s ($a$=1.34 s) for the F (U) state.  
({\bf d}) 
Typical force and extension traces (upper and middle) obtained by  
applying a force oscillation protocol with amplitude $A=0.7$~pN and 
frequency $\nu_{\rm os}=0.4$~Hz around the coexistence force 
$f_{\rm c}\simeq 14.5$~pN. In the lower panel, we show the measured power 
spectrum, $S(\nu)$, calculated by Fast Fourier Transform with window size 
$N=2^{17}$ of $X(t)$ shown in the middle panel. 
The sampling rate of the instrument 
is 1~kHz. The red area is the output signal (OS, Eq.\ref{os}) and the blue 
vertical bar represents the background noise (BN, Eq.\ref{in}).
}
\label{fig1}
\end{figure*}

In this work, we 
use optical tweezers to investigate SR in single DNA hairpins driven by 
oscillatory mechanical forces. The high chemical stability of DNA makes 
DNA hairpins excellent models to 
investigate SR at the single-molecule level. When force oscillates around 
the average unfolding force, thermally 
activated hopping kinetics between the folded (F) and unfolded (U) states 
synchronizes with the frequency of the external driving force, leading to 
SR.  SR can be measured by recording 
the oscillations produced in the molecular extension, relative to the 
magnitude of the noise produced by the thermal forces.
 Our aim in this work is to perform  a systematic study of SR in 
single-molecules exhibiting bistable 
dynamics, rather than using SR as a useful tool to determine the kinetic 
properties of DNA hairpins.  In fact, these can be estimated by 
using other much less time-consuming methods (e.g. by 
directly analyzing hopping traces). Yet, we also carry out SR studies in 
short hairpins that show how SR might prove useful to enhance the detection 
of conformational transitions of low SNR.

The paper is organized as follows.  In Section II, our experimental set
up is explained. Our main SR results in DNA hairpins are presented in
Section III, and the influence of the experimental conditions
(i.e. dsDNA handle length and trap stiffness) is investigated in Section
IV.  We compare different SR quantifiers in Section V and in section VI
we describe the related phenomenon of resonant activation.  Finally in
Section VII, we purposely designed short DNA sequences to increase the
noise of the signal to test whether SR can still be used to identify the
hopping rate.  In the last section, we summarize our conclusions, and
discuss situations where SR might be a useful technique.

\section{Experimental setup and Hopping Experiments}

In Fig.~1a, we show a schematic illustration of our experimental setup (left) 
and the DNA sequence of hairpin H1 that we investigated (upper right). 
The DNA hairpin is tethered between two short dsDNA handles (29 bp) that 
are linked to micron-size beads \cite{Nuria}. One bead is captured in the 
optical trap whereas the other is immobilized at the tip of a glass pipette 
\cite{footnote1}.   
By moving the position of the optical trap relative to the pipette, 
a force is exerted at the extremities of the hairpin.  

In a pulling experiment, the optical trap is moved away from the pipette 
and mechanical force is applied to the ends of the DNA construct (DNA 
hairpin plus DNA handles) until the value of the force at which the hairpin 
unfolds is reached. In the reverse process, the trap approaches the pipette 
and the force is relaxed until the hairpin refolds. In this experiment, the 
force exerted upon the system, $f$,  is recorded as a function of the relative 
trap-pipette distance giving the so-called force-distance curve 
(Fig. 1a, lower right).  Around the co-existence force, 
$f_{\rm c}\simeq14.5$~pN, the hairpin hops between the F and U states for 
sufficiently low pulling speeds. 

Hopping experiments can be done in two different modes: constant 
force mode (CFM) and passive mode (PM) \cite{exp2,exp3}. In the CFM, the 
force applied to the DNA construct is maintained at a preset value by 
moving the optical trap  through force-feedback control (Fig. 1b, upper). 
 The folding and unfolding 
transitions of the DNA hairpin are followed by recording the trap position, 
$X(t)$. In contrast to the CFM, the PM is operated by leaving the position 
of the optical trap stationary without any feedback. The bead passively 
moves in the trap in response to changes in the extension of the DNA 
construct (Fig. 1b, lower). When the hairpin unfolds, the trapped bead moves 
toward the trap center 
and the force decreases; when the hairpin folds, the trapped bead is 
pulled away from 
the trap center and the force increases. The folding and unfolding 
transitions of the 
DNA hairpin are followed by recording the force, $f(t)$. In both cases 
(CFM and PM), the kinetic rates of hopping can be measured from the 
 residence times  
of the trace ($X(t)$ in the CFM and $f(t)$ in the PM). Fig. 1b 
shows hopping traces measured in the CFM and PM at the co-existence 
force, $f_{\rm c}\simeq 14.5$~pN, where the hairpin hops between the F and 
U states populating them with equal probability (i.e. it spends equal time 
in both states).

In this work, we focused on the experiments at controlled force, 
rather than at fixed trap position. Both the hopping and the 
oscillation experiments (described below) were carried out using 
the force feedback control. The reason is that the controlled force 
experiments avoid undesirable drift effects in force that strongly 
affect the kinetics of the hairpin (see Methods). Therefore we mainly 
carried out the experiments in the CFM by recording the position of the 
trap, $X(t)$. This signal exhibits dichotomic motion between the two 
distinct levels of extension (Fig. 1b, upper left).  The difference 
between the two levels (short extension, folded; long extension, 
unfolded) reflects the release in extension ($\simeq 18$ nm) of the 
44~nucleotides of hairpin H1. From $X(t)$ we can extract the residence 
time distribution at each state that shows the  
exponential form characteristic of first-order decay processes (Fig. 1c). 
The fit of the 
time distribution to an exponential function allow us to get the average 
residence time. The force-dependent kinetic rates (equal to the 
inverse of the mean lifetimes), 
$k_{\rm FU}$ and $k_{\rm UF}$, were measured at the co-existence force, 
$f_{\rm c}=14.5\pm0.3$~pN, giving 
$k_{\rm c}=k_{\rm FU}^{\rm c}=k_{\rm UF}^{\rm c}\simeq$ $0.66\pm0.04$~s$^{-1}$  
(Table S0 in SI).

\section{SR experiments}

To induce the SR phenomenon, we applied an oscillating force, $f(t)$, to the
DNA hairpin using the force feedback protocol, where 
$f(t)=f_{\rm c}+f_{\rm os}(t)$.  For $f_{\rm os}(t)$ we chose a square-wave 
signal of amplitude, $A$, and frequency, $\nu_{\rm os}=1/T_{\rm os}$, where 
$T_{\rm os}$ is the oscillation period (Fig.~1d, upper). The four distinct 
levels of extension observed (Fig. 1d, middle) correspond to the 
molecular extensions of the hairpin in the F and U states at the two force 
values, $f=f_{\rm c}+A$ and $f=f_{\rm c}-A$. The power spectral density, 
$S(\nu)$, is defined as the Fourier transform of the stationary 
correlation function of the signal $X(t)$:
\begin{equation}
S(\nu)= \int_{-\infty}^{+\infty}{\rm d}t \bra X(t)X(0)\ket 
{\rm e}^{-i2\pi\nu t},
\label{power}    
\end{equation}  
where $\bra\cdot\ket$ denotes a time-average over the signal. As shown 
in Fig.~1d (lower), $S(\nu)$ can be described as the superposition of a 
background power spectral density, $S_N(\nu)$, and a structure of delta 
spikes centered at $\nu_n=(2n+1)\nu_{\rm os}$ ($n=0,1,2,\cdots$). In order 
to extract the signal from the background noise, we define the output 
signal (OS), the background noise (BN) and the signal-to-noise ratio (SNR) 
as \cite{hanggi},
\begin{eqnarray}
&&{\rm OS}= \lim_{\Delta \nu\to 0} 
\int_{\nu_{\rm os}-\Delta \nu}^{\nu_{\rm os}+\Delta\nu} S(\nu) d\nu.
\label{os}\\
&&{\rm BN}=S_N(\nu_{\rm os}).
\label{in}\\
&&{\rm SNR}= \frac{{\rm OS}}{{\rm BN}}
=\frac{1}{S_N(\nu_{\rm os})}\lim_{\Delta \nu\to 0}
\int_{\nu_{\rm os}-\Delta \nu}^{\nu_{\rm os}+\Delta\nu} S(\nu) d\nu.
\label{snr} 
\end{eqnarray}
The SNR defined in Eq.(\ref{snr}) is equal to the ratio of the spectral 
power of the signal at the frequency $\nu_{\rm os}$ (OS), to the noise-floor 
spectral density measured in the presence of the oscillation (BN) and has 
dimensions of Hz.  Fig. 1d (lower) illustrates how we measured 
the OS (red area) and the BN (blue vertical bar) from the spectral density. 
Other equivalent definitions of the SNR \cite{NamWie89} are the 
dimensionless ratio between the power in the output signal  
(Eq.(\ref{os})) 
and the total input noise power 
delivered by the noise (proportional to the integral of 
background spectral density $S_N(\nu)$ 
over all $\nu$). Because the total input noise 
power only depends weakly on $\nu_{\rm os}$, we 
can take the OS, Eq.(\ref{os}), as 
another indicator of the SR phenomenon. Indeed, both indicators OS and SNR 
are equally valid to identify resonant behavior, even though the peak is 
often more visible in the latter (see below) \cite{Stocks}.

\begin{figure}
\begin{center}
\includegraphics[width=8cm]{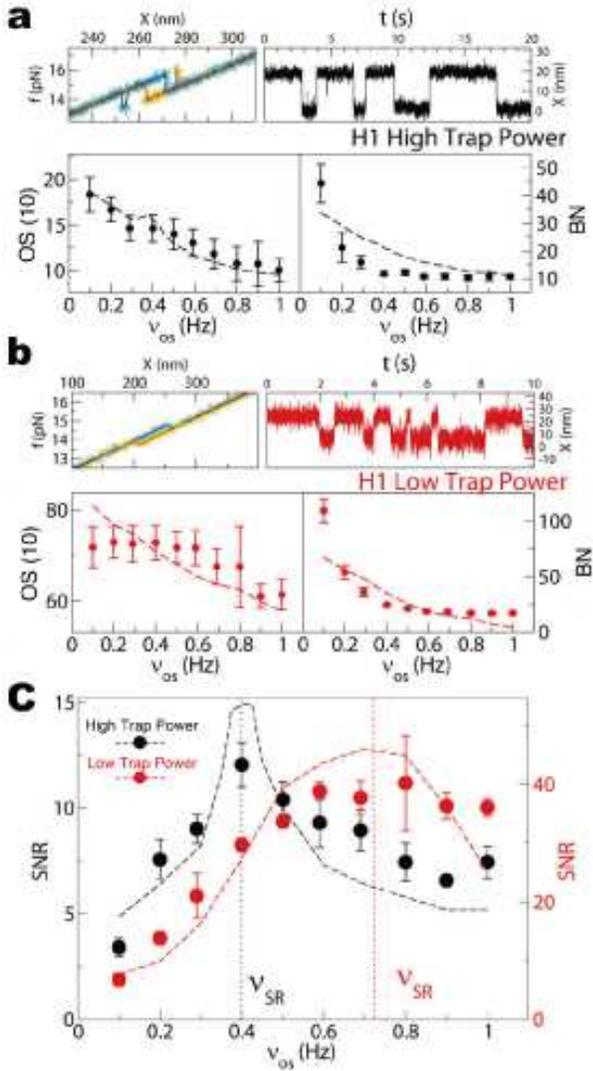}
\end{center}
\caption{ {\bf SR experiments for hairpin H1 at different trap stiffness.}
Pulling cycle (unfolding, blue; refolding, red), 
hopping trace, OS and BN for H1  with amplitude $A=0.7$~pN  
at high trap stiffness, $\kappa_{\rm trap}=70$ pN/$\mu$m ({\bf a}), 
and at low trap stiffness, $\kappa_{\rm trap}=24$ pN/$\mu$m  ({\bf b}). 
Results averaged over 5-10 molecules. 
(Note: the force rips shown in force-distance curves
should drop vertically without any finite stiffness correction. The
  finite slope correction shown in Fig. 2b (top, left) is due to 
low-bandwith filtering of data).  ({\bf c})  SNR at high trap stiffness   
(low trap power) depicted in black (red).   
Units: OS (nm$^2$), BN (nm$^2$/Hz), SNR (Hz). Simulation results are 
shown as dashed lines (Figs.~S4 and S7 in~SI).   
The error bars represent the standard error over different molecules. 
}
\label{fig2}
\end{figure}

\begin{figure}
\begin{center}
\includegraphics[width=8cm]{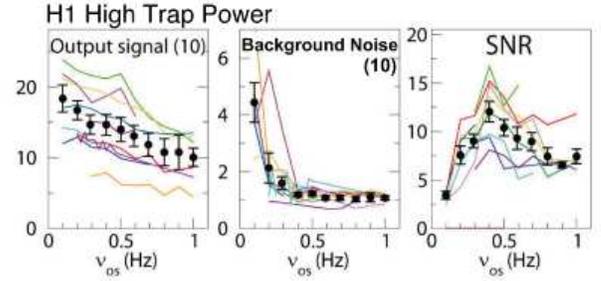}
\end{center}
\caption{ {\bf Molecular variability of the measured responses (H1).} 
Results for the OS, BN and SNR for 10 different molecules at high trap 
power $\kappa_{\rm trap}=70$ pN/$\mu$m. Units: OS (nm$^2$), BN (nm$^2$/Hz), 
SNR (Hz). The error bars represent the standard error over different molecules.
}
\label{fig3}
\end{figure}

For the hairpin H1 at high trap power and trap
stiffness $\kappa_{\rm trap}\simeq 70$~pN/$\mu$m, the resulting OS
and BN as a function of $\nu_{\rm os}$ are depicted in Fig. 2a
(lower), while Fig.~2c shows the SNR. In contrast to the OS, the
presence of a peak around $\nu_{\rm os}=0.4\pm0.05$~Hz is apparent for
the SNR. This value is close to that predicted by the matching
condition, $\nu_{\rm SR}=k_c/2$, which states that the SNR is maximum when 
the average hopping
time of the hairpin ($1/k_{\rm c} = 1.56$~s) is
equal to half the period of the forcing oscillation ($1/2{\nu_{\rm
    os}}$=1.25~s) \cite{hanggi,pre,snr1,snr2}.  
This shows that SR in single-molecule hopping experiments approximately 
fulfills the matching condition as has been observed in other bistable 
systems.

The OS and the SNR can be calculated theoretically as a function of 
the oscillation frequency 
for a Brownian particle in a continuous double-well potential \cite{Stocks,pre,
casado}. In this model, the OS and the SNR exhibit a soft and a 
sharp peak, respectively, only when SR is induced at large enough forcing  
amplitudes \cite{Stocks}. 
These large forcing amplitudes correspond to a non-linear regime of the 
system, in which the shape of the double-well potential is so deformed 
that the barrier separating the wells vanishes at the maximum elongation 
of the oscillation. In our experiments, we applied a large oscillation 
amplitude ($A$=0.7~pN). Note that the region 
of coexistence between the F and U states spans less than 3~pN in Fig.~1a 
(lower right). Thus an extra-force of 0.7~pN strongly alters the barrier 
and the relative free energy between states F and U. 
Our experimental results agree with the 
theoretical predictions by Stocks \cite{Stocks} obtained in the 
non-linear response regime. We performed a numerical simulation of an 
overdamped particle moving in a double-well potential with parameters that 
fit the experimentally measured molecular free energy landscape (Sec. IV 
in SI). Despite its 
simplicity, the model qualitatively reproduced the experimental 
results for the OS, BN and SNR  (dashed lines in Fig.~2c). 

In order to see 
what happens for lower oscillation amplitudes, we explored the response of 
hairpin H1 to an oscillating force of lower amplitude, $A=0.2$~pN. 
A very soft peak and a gentle maximum in the OS and the SNR can be seen 
around 0.4~Hz (Fig. S1 in SI) in agreement with the results previously 
obtained for the 
higher amplitude, $A=0.7$ pN  (Fig. 2). 
However, the peak for $A=0.2$~pN is much less clear than the peak 
for $A=0.7$~pN, showing the importance of using oscillation amplitudes 
beyond the linear-response regime ($AX^{\dagger\dagger}/k_BT \ll 1$, 
where $X^{\dagger\dagger}$ is the characteristic distance separating the 
folded or unfolded states from the transition state. See also 
Sec. III in SI for SR behavior in the linear response regime). 

A characteristic feature of SR experiments at the single-molecule
level is the large variability observed in the measured response from
different molecules. Fig.~3 shows the OS, BN and SNR for 10 different
molecules.  Larger variability is observed for the OS as
compared to the BN. This might be due to non-linear effects which are
sensitive to small differences in the experimental setup (e.g. 
 tether misalignment with respect to the
pulling direction, variations in the size of the bead and the
trap stiffness, etc.).


\section{Influence of trap stiffness and length of the handles}

An important issue in single-molecule experiments concerns the
influence of transducing effects induced by the experimental setup
(e.g. trap stiffness and length of the handles) on the measured kinetics.
Recent studies \cite{exp2,exp3,Nuria} show that the kinetic rates are
only moderately affected (within one order of magnitude) when changing
the length of the handles one thousand-fold or the trap stiffness ten-fold. 
Besides, numerical simulations carried out in Ref. \cite{exp3}
show that kinetic rates for hairpins measured with handles and trap
always remain close and converge to the intrinsic rate (i.e. the rate
measured without handles and trap) in the limit of very compliant
linkers.  To inquire the influence of the experimental design on the
kinetics of hairpin H1, SR was further investigated by varying
conditions of the experimental setup such as 1) the stiffness of the
optical trap and 2) the length of the handles.  We observed how both
effects changed the intrinsic noise of the system (Figs.~2b, 2c and
4).  In the first case, when the trap stiffness, $\kappa_{\rm trap}$,
was decreased from $70$ pN/$\mu$m to $24$ pN/$\mu$m (Fig.~2b), the maximum
peak in the SNR was shifted to higher frequencies (from 0.4~Hz to
$\simeq0.8$~Hz) and became less clear (Fig. 2c, red curve).  The
effect of the trap stiffness on SR was evaluated by using the
numerical simulation (Sec. IV in SI), finding good agreement
between experiments and simulations (Figs. 2b and 2c).

In the second case, if we increase by twenty-fold the length of the handles  
(528~bp and 874~bp at each flanking side) keeping the trap stiffness 
constant, $\kappa_{\rm trap}=70$~pN/$\mu$m, we find that the resonance 
frequency shifts to a larger value for the long handles (Fig. 4). For the 
long handle construct, the matching condition is verified 
($\nu_{\rm SR}=2$ Hz) and  $k_{\rm c}\simeq 4$~${\rm s}^{-1}$ as obtained 
from hopping experiments \cite{Nuria}. 

The dependence of the resonance frequency measured from SR, $\nu_{\rm SR}$, 
on the trap stiffness and the length of the handles was similar to that 
reported for the hopping rate measured in the hopping experiments at the 
co-existence force \cite{exp2,exp3,Nuria}. In both cases, the soft trap 
stiffness or the larger compliance of the long handles contributes to 
increase the hopping rate, supporting the conclusions of Ref.~\cite{Nuria}. 
Interestingly enough, the quality of the resonant peak worsens as the 
trap stiffness decreases but not as the linker becomes softer, showing that 
the quality of the SR peak is only dependent on the combined effective 
stiffness of bead and handles ($\kappa^{-1}_{\rm eff} = \kappa^{-1}_{\rm trap} 
+ \kappa^{-1}_{\rm handle} \simeq \kappa^{-1}_{\rm trap}$) which is 
approximately equal to the trap stiffness in our experimental conditions.

\begin{figure}
\begin{center}
\includegraphics[width=8cm]{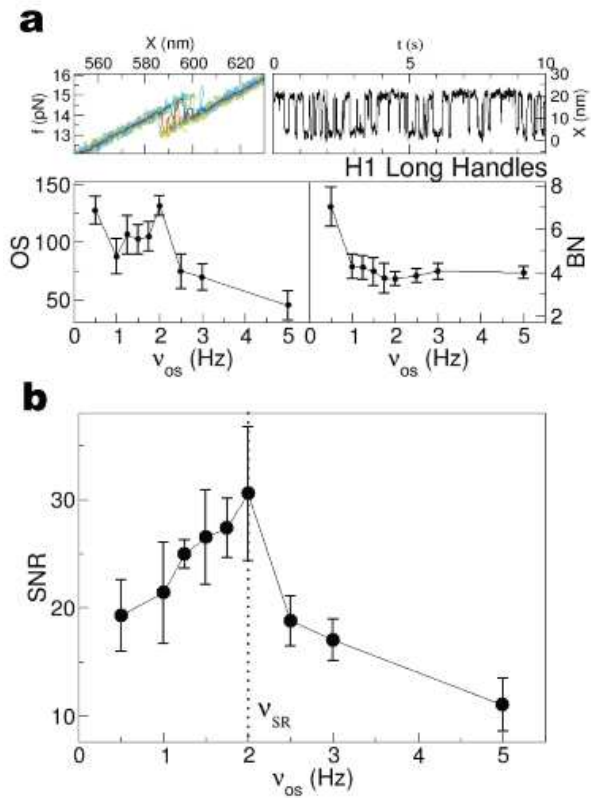}
\end{center}
\caption{ {\bf SR experiments for hairpin H1 with long DNA handles.} 
({\bf a}) Pulling cycle (unfolding, blue; refolding, red), hopping trace, 
OS and BN depicted in violet. 
({\bf b}) The resulting SNR in the case of high trap stiffness  
$\kappa_{\rm  trap}=70$ pN/$\mu$m and the amplitude $A=0.7$~pN. 
Results averaged over 5 molecules. Units: OS (nm$^2$), BN (nm$^2$/Hz), SNR (Hz).
The error bars represent the standard errors over different molecules.
}
\label{fig4}
\end{figure}

\begin{figure*}
\begin{center}
\includegraphics[width=16cm]{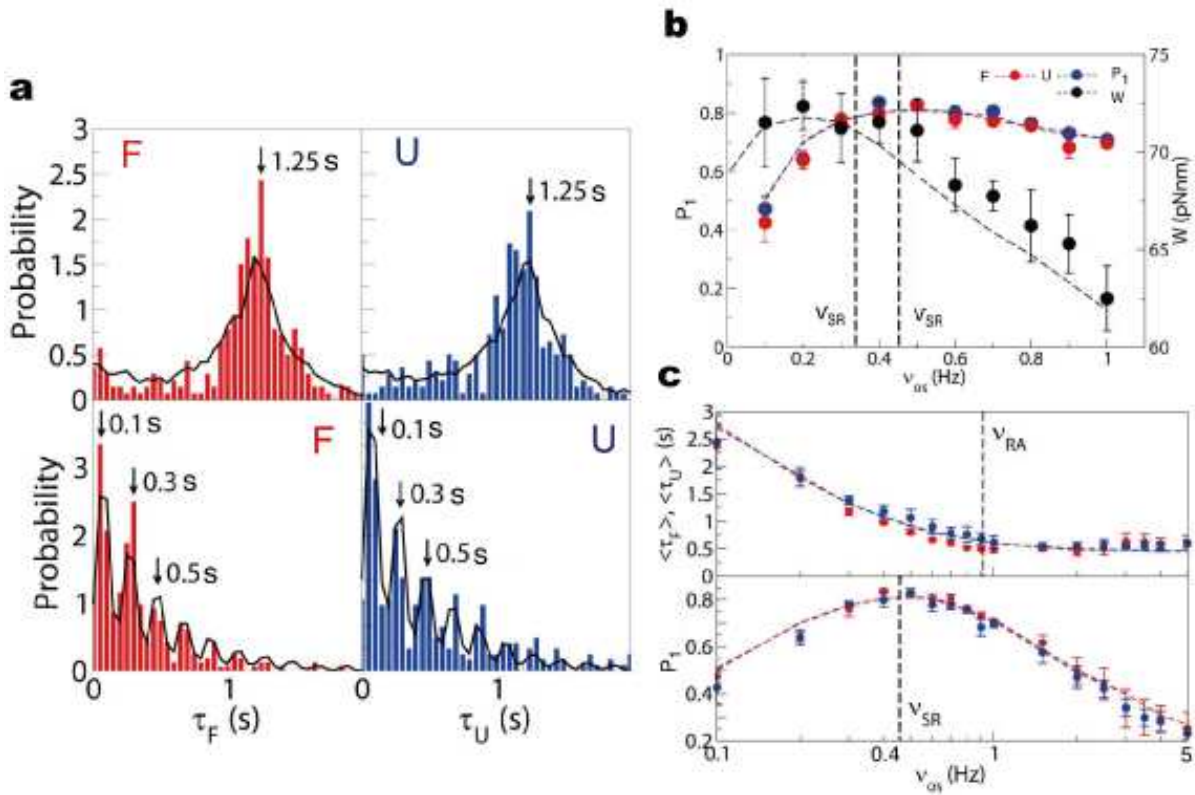}
\end{center}
\caption{ {\bf Other SR quantifiers in hairpin H1.} ({\bf a}) Residence time
distributions for the F (red) and U (blue) states at $\nu_{\rm
  os}=0.4$~Hz (upper) and $\nu_{\rm os}=5$~Hz (lower). The continuous
black lines are the results of the simulations (Sec. IV in SI). ({\bf b})  
$P_1$ in the F (red) and U (blue) and average work $W$ (black) as 
a function of input frequency. Simulation results are 
shown by the dashed lines (Sec. IV in SI). 
Vertical dotted line shows 
the expected resonance frequency. The maximum in $P_1$ is broad 
around 0.3-0.6 Hz, while that in $W$ is also broad with a maximum found 
at a lower frequency in the range 0.1-0.4 Hz. 
({\bf c}) Average residence times (upper panel) and $P_1$ (lower panel) 
in the F (red) and  U (blue) states as a function of input frequency. 
Note that the frequency range is larger than that shown in b). 
Simulation results are shown as dashed continuous lines 
(Sec. IV in SI). The vertical lines show the two frequencies characteristic 
of stochastic resonance ($\nu_{\rm SR}$) and resonant activation 
($\nu_{\rm RA}$). Statistics: 8 molecules, 3-5 minutes traces and 
300-600 hopping transitions at each input frequency. 
The error bars represent the standard errors over different molecules. 
}
\label{fig5}
\end{figure*}

\section{Other SR quantifiers}

Next we investigated other representative SR quantifiers. These are:
the fraction $P_1$ of transitions that occur every half-period of
the oscillation \cite{p1,p1-2, schmitt}; and the average dissipated
work, $W$~\cite{work,ciliberto}.  To extract $P_1$, we measured the
residence time distributions, $P(\tau_{\rm F})$ and $P(\tau_{\rm U})$,
of the F and U states in the presence of the oscillating force.  The
distributions are shown in Fig.~5a for hairpin H1 in the cases
$\nu_{\rm os}=0.4$~Hz (upper) and $\nu_{\rm os}=5$~Hz (lower) with
$A$=0.7~pN. Unlike the distributions shown in Fig.~1c, $P(\tau_{\rm
  F})$ ($P(\tau_{\rm U})$) is not monotonically decreasing with
$\tau_{\rm F}$ ($\tau_{\rm U}$) and exhibits spikes corresponding to
higher harmonics for $\tau_{\rm F}= T_{\rm os}(1+2n)/2$ ($\tau_{\rm
  U}= T_{\rm os}(1+2n)/2$) where $n=0,1,2,\cdots$. A few harmonic
frequencies are shown as vertical arrows in Fig.~5a. In particular
when $\nu_{\rm os}$ is close to the resonance frequency, the shape of
the residence time distribution strongly deviates from an exponential
and a broad peak appears around the fundamental mode, $\tau_{\rm
  F}=T_{\rm os}/2$ ($\tau_{\rm U}=T_{\rm os}/2$) (Fig.~5a, top). In
contrast, many peaks appear in the residence time distribution when
$\nu_{\rm os} \gg \nu_{\rm SR}$ (Fig.~5a, lower).

$P_1$ can be extracted from the area of the residence 
time distribution around the peak located at the fundamental mode, 
$\tau_{\rm F}=T_{\rm os}/2$ ($\tau_{\rm U}=T_{\rm os}/2$).  
Let $\lbrace \tau_i ;i=1,\cdots,N \rbrace$ be the 
series of $N$ residence times measured in the presence of the 
oscillating force.  By counting the number, $n$, of $\tau_i$ that satisfy  
the condition $T_{\rm os}/2-T_{\rm os}/4 \le \tau_i \le T_{\rm
  os}/2+T_{\rm os}/4$, we define
\begin{equation}
P_1=\frac{n}{N}. 
\label{pp1}
\end{equation}
$P_1$ takes a large value if the residence time of the hairpin is equal 
to half the period of the oscillating force. This means that a large 
fraction of hopping transitions occur when the oscillating force changes 
sign. Therefore, the value of $P_1$ has a maximum when SR is induced, 
because the transitions between the two states are synchronized 
with the oscillating force ($P_1$ is a {\it bona fide} SR quantifier 
\cite{p1}. See also Sec. III in SI).    
The results obtained for $P_1$ in hairpin H1 are shown in Fig. 5b. $P_1$ 
exhibits a broad maximum around the resonance value 
$\nu_{\rm SR}=k_{\rm c}/2=0.4$~Hz. The broadness of the peak is in contrast 
to the narrower peak observed in the SNR (Fig. 2c). These results are  
consistent with analytical calculations  \cite{p1,hanggi}.  

For the average cyclic work done by an oscillating force,  
we define \cite{Mossa09} 
\begin{equation}
W= -\langle\oint X df\rangle=\langle\oint f dX\rangle, 
\label{work}
\end{equation} 
where the brackets stand for statistical averages over traces. Because
$W$ takes a large value when the folding/unfolding of the hairpin is
synchronized with the oscillating force, it is a useful SR quantifier as
well \cite{ciliberto,jung}. 
In fact, the larger the synchronisation between
transitions of the hairpin and oscillations in the force, the larger the
work done by the optical trap on the molecule.  
Results for $W$ are
shown in Fig. 5b. In contrast to SNR but similarly to $P_1$, the maximum
in $W$ is broad. Finally, we compared our experimental results with the
predictions obtained from the numerical simulations in
the continuous double-well potential whose parameters are 
the same as those used in Fig. 2  (Sec. IV in SI).  Figs.~5a 
and 5b show a good agreement between experiments and simulations. 
Although both $P_1$ and $W$ show broad maxima 
as a function of $\nu_{\rm os}$, they are not coincident, the maximum 
for the work is found at a lower frequency as compared to $P_1$. As 
pointed out in the introduction, the precise value of the resonance frequency 
depends on the quantifier specially when the quality of the resonant 
peak is low.

\begin{figure*}
\begin{center}
\includegraphics[width=16cm]{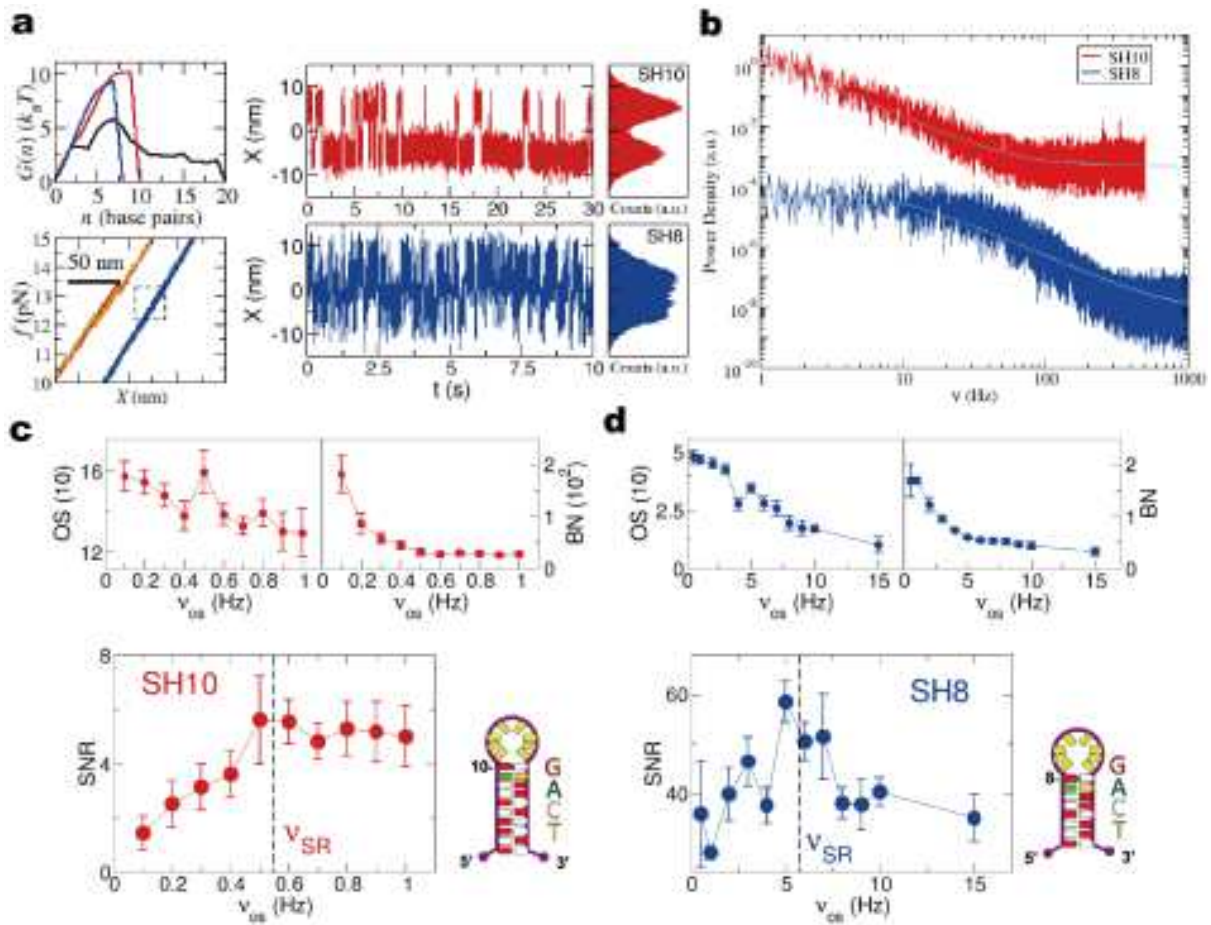}
\end{center}
\caption{ {\bf SR experiments in shorter hairpins.} ({\bf a}) Free
  energy landscapes (upper left), force-distance curves (lower left) and hopping
  traces in the CFM for SH10 (red), SH8 (blue) and H1
  (black). Measurements were carried out with a low trap power
  $\kappa_{\rm trap}\simeq 32$~pN/$\mu$m for SH10, 17~pN/$\mu$m for
  SH8, respectively. For sake of simplicity, all pulling curves in the
  lower left panel are 
shown parallel with equal average slopes. The dashed-squared
    region for SH8 curves indicates the region where unfolding/folding
    transitions occur.  Distributions of trap position, $X$, show
  clear transitions for SH10 but not for SH8. ({\bf b}) Power spectrum
  for SH10 and SH8.  Cyan curves are fits to a sum of two
  Lorentzians (see text for details). Colors as in Fig. 6a. 
({\bf c, d}) OS, BN and SNR for
  SH10 and SH8. The amplitudes of oscillation force are
  $A=0.5$ pN for SH10 and $A=0.15$ pN for SH8, respectively.  Colors as 
in Fig. 6a. Statistics
  (SH10, SH8): Molecules (5, 7); Duration of hopping traces (4, 2)
  minutes; Average number of hopping transitions (250, 1200) at each
  input frequency. Units: OS (nm$^2$), BN (nm$^2$/Hz), SNR (Hz).  The
  error bars represent the standard errors over different molecules.
}
\label{fig6}
\end{figure*}

\section{Resonant activation}
In stochastic systems driven by oscillating forces, it is customary to
distinguish two effects: stochastic resonance (SR) and resonant
activation (RA). SR stands for the optimization of the response of the
system (i.e. the output signal) whereas RA stands for the optimization
of kinetics (i.e., maximization of the number of hopping transitions
per second). SR and RA are different phenomena related to barrier
crossing dynamics along temporally modulated energy landscapes
\cite{schmitt}. RA is induced when the mean residence times of the
states of the system are minimized with respect to the frequency of
the oscillating force, $\nu_{\rm RA}$.  The values of $\nu_{\rm SR}$
and $\nu_{\rm RA}$ are often not the same, the latter being typically
larger than the former. Fig.~5c (top) shows the mean residence times,
$\bra\tau_F\ket$ and $\bra\tau_U\ket$, for hairpin H1 measured in the
range $0.1$ Hz $\le \nu_{\rm os} \le 5$~Hz. Only at higher frequencies
(between 1~Hz and 2~Hz), the graph suggests a very shallow minimum for
the residence times.  Therefore we are capable of observing both the
SR and RA phenomena in the single-molecule experiments. The
  experimental results also agree with the numerical simulations
  (Fig. 5c, dashed lines). Similar behavior has been reported in the
experiments with a colloidal particle in a double-well potential
generated by optical tweezers \cite{schmitt}.

\section{SR in shorter hairpins}

SR might be used to detect the transitions in cases where the hoppings of
a hairpin is hard to be discriminated. These correspond to
cases in which the hopping signal (extension jumps) are on the
same order of the standard deviation of noise fluctuations. To
investigate this problem, we designed two short hairpins (SH10 and SH8)
having only 10 and 8 base pairs along the stem, respectively (sequences 
shown in Figs. 6c and 6d). The molecular free energy
landscapes were calculated for the two sequences at the 
theoretically predicted co-existence forces using the 
  nearest-neighbour model for DNA (Fig. 6a, upper left) 
  \cite{design,Mossa09b}. As the length of the stem decreases, the
  landscapes show progressively lower co-existence force values,
  molecular extensions and kinetic barriers. Measurements for SH10 and
  SH8 were taken at low trap stiffnesses to decrease the hopping signal 
($\kappa_{\rm trap}\simeq
  32$~pN/$\mu$m and 17~pN/$\mu$m, respectively). Pulling curves and
  hopping traces in the CFM are also shown in Fig.~6a (lower left). While the
  transitions are still visible for SH10, these are hardly
  discriminated for SH8. This is also apparent from the dwell
  distributions on trap position, $X$, shown in Fig. 6a (right). Measured jumps
  in the molecular extension upon unfolding/folding are equal to
  $10.5\pm0.5$~nm and $7.0\pm0.5$ nm for SH10 and SH8, respectively.

Fig. 6b shows the power spectra of $X(t)$. Whereas SH10 can be fit reasonably 
well to a sum of two Lorentzians with two characteristic corner frequencies 
(0.64$\pm$0.02~Hz and 2.4$\pm$0.3~kHz), the quality of the fit considerably 
worsens for SH8 (\mbox{$\simeq$ 9.8~Hz} and \mbox{$\simeq$ 15.6~kHz}). 
The low frequency (in the range of Hz) in the power spectra corresponds to the 
hopping kinetics of the hairpin whereas the high frequency (in the range of 
kHz) corresponds to the random motion of the optical trap caused by the 
force feedback. Because the noise in the trap position, $X$, introduced 
by the force feedback protocol is not of thermal origin, the power 
spectra measured in the CFM should not necessarily be fit to a sum of two 
Lorentzians. This is specially acute for SH8  where the feedback loop 
cannot follow the fast hopping transitions. 

Once the hopping properties 
of the hairpins were characterized, we then carried out the oscillating 
experiments for hairpins SH10 and  SH8 around the co-existence force. 
The results we 
obtained for SH10 are similar to those reported for hairpin H1 at low 
trap power shown in Fig. 2c. For SH10 the peak in the SNR around 
$\nu_{\rm SR}$=0.5~Hz is 
close to $k_{\rm c}$/2 where $k_{\rm c}$ was measured to be 
0.43$\pm$0.07s~$^{-1}$ 
from the hopping traces for $X(t)$. More interesting is 
the case of hairpin SH8 where the co-existence force can still be located, 
but the hopping signal is blurred by the fluctuations. In Fig. 6d, we can 
see that the OS and the SNR exhibit a maximum around $\nu_{\rm SR}=5\pm1$~Hz 
for SH8 
which gives $k_{\rm c}\simeq 10\pm2$~${\rm s}^{-1}$ according to the 
matching condition.  This value agrees with the value of \mbox{$\simeq 9.8$~Hz} 
obtained from the Lorentzian fit to the power spectrum. 
As an additional test, we have implemented a Hidden Markov Model (HMM) with 
the forward-backward feedback algorithm as described in Ref.~\cite{hmm} to 
extract the kinetic rates of SH8 from the hopping trace, 
$X(t)$. By applying the HMM to the hopping traces of SH8, we obtained a value 
of $k_{\rm c}=9.4\pm 0.5$ s$^{-1}$ (7 molecules), which 
confirms the results obtained with SR and Lorentzian fit to the spectral 
density. 

Thus, SR offers an alternative method to estimate the hopping rate of SH8. 
Indeed, the two states (F and U) cannot be easily detected from the hopping 
trace and the residence time analysis done for hairpin H1 (Fig.~1c) is 
difficult to 
implement. In this case SR confirms the value of the hopping frequency 
initially obtained from a poor Lorentzian fit of the power spectrum.

\begin{table*}
\begin{center}
\begin{tabular}{|c|c|c|c|c|c|}
\multicolumn{3}{c}{\bf Comparison between $\nu_{\rm SR}$  (Hz) 
and $k_{\rm c}$}(s$^{-1}$) \\
\hline 
   & $\nu_{\rm SR}$ from SNR & $\nu_{\rm SR}$ from OS   &
$\nu_{\rm SR}$ from $P_1$ &  
$\nu_{\rm SR}$ from $W$
& $k_{\rm c}/2$  \\
\hline
H1$^{\rm a,c}$   & 0.40$\pm$0.02 ($n=10$) & 0.45$\pm$0.03 ($n=10$) & 0.45$\pm$0.04$^{\rm e}$, 0.48$\pm$0.08$^{\rm f}$ ($n=8$) 
 & 0.33$\pm$0.07 ($n=8$) & 0.33$\pm$0.02 ($n=12$)\\ 
\hline
H1$^{\rm a,d}$ & 0.72$\pm$0.08 ($n=5$) & 0.50$\pm$0.06 ($n=5$) &- &- & 0.53$\pm$0.07 ($n=8$) \\
\hline
H1$^{\rm b,c}$  & 2.0$\pm$0.2 ($n=5$) & 1.9$\pm$0.1 ($n=5$)  & -& -& 2.2$\pm$0.3 ($n=5$) \\
\hline
SH10 & 0.54$\pm$0.02 ($n=5$) & 0.58$\pm$0.04 ($n=5$) &- &- & 0.43$\pm$0.07 ($n=4$) \\
\hline
SH8 & 5.7$\pm$0.4 ($n=7$) & 5.3$\pm$0.2 ($n=7$) & -& -& 4.7$\pm$0.3$^{\rm g}$ ($n=7$) \\
\hline
\end{tabular}
\caption{Resonance frequency, $\nu_{\rm SR}$, obtained from SNR, OS, 
P$_1$ and $W$  vs. 
hopping rate, $k_{\rm c}$, at the co-existence force (see Sec. I in SI 
for  $k_{\rm c}$).   $\nu_{\rm SR}$ was chosen as the peak value of each 
SR quantifier for each molecule. 
(a. Short handles. 
b. Long handles.  
c. High power trap ($\kappa=70$ pN/$\mu$m). 
d. Low power trap ($\kappa=24$ pN/$\mu$m). 
e. Folded state. 
f. Unfolded state. 
g. Rate determined using a Hidden Markov Model (HMM). 
Note that $n$ is the number of molecules analyzed.)   
}
\end{center}
\vspace{-0.6cm}
\label{table1}
\end{table*}

\section{Conclusion}

We carried out SR experiments in single DNA hairpins subject to an
oscillatory mechanical force of varying frequency. Our aim was to 
investigate how a molecule exhibiting bistability (i.e. hopping between 
the folded and unfolded conformations) responds to an applied oscillating
force. In SR the response gets amplified at 
frequencies close to the characteristic hopping frequency of the 
hairpin. By measuring the power spectral density of the molecular 
extension, we carried out a detailed investigation of the 
frequency dependence of the output signal (OS, Eq. (2)), 
the background noise (BN, Eq. (3)) and  
the signal-to-noise ratio (SNR, Eq. (4)) in the 20bp hairpin H1 which exhibits 
dichotomous hopping behavior. We then extended our 
research by exploring how several parameters of the experimental setup 
such as trap stiffness, length of the handles, oscillating amplitude 
and size of the hairpin influence the resonance behavior. From the measured  
traces, we also analyzed a few other SR  
quantifiers such as the  
number of folding and unfolding transitions occurring 
every half-period of the oscillation ($P_1$, Eq. (5)), 
the average mechanical work per period of the oscillation 
($W$, Eq. (6)) and the mean  
residence times  
in the unfolded and folded states ($\bra\tau_{\rm U}\ket$ and 
$\bra\tau_{\rm F}\ket$).  The mean residence times 
describe a mechanism slightly different from SR that has  
been termed resonant activation (RA). 
Overall, we find that the SNR and the  
other SR quantifiers (such as OS, $P_1$, $W$) exhibit a peak at a 
frequency close to that determined by the resonance matching  
condition. Among all quantifiers only the SNR and the OS 
tend to show a modest amplification of the response, the SNR
  showing a higher quality peak. 
Our results are summarized in Table I. 
Moreover, our experimental results are well predicted by numerical  
simulations of an overdamped particle in a 
double-well potential 
reproducing the measured molecular free energy landscape of the 
hairpin (Sec. IV in SI). 
Finally, our experimental findings also agree with theoretical results  
\cite{Stocks} that show a modest gain 
in the response of noisy systems driven by  
oscillating forces. 

A unique aspect of our work is the investigation of SR in small 
 systems in conditions of weak thermodynamic stability (folding free 
energies of a few $k_{\rm B}T$ units) not far from noise level 
($k_{\rm B}T$). This has a primary consequence: the proper control 
  parameter in our experiments does not appear to be the noise
  intensity. In fact, by changing noise intensity (e.g. by tuning
  temperature or denaturant concentration), we also modify the
  structural properties of the molecule in a non-controlled way (i.e. by
  changing its thermodynamic stability or free energy of formation).
  Our work circumvents this problem by using the frequency of the
  external driving force as control parameter. Simple as this choice may
  seem only a few theoretical and experimental works have addressed
  it in the past. From this perspective, our study should stimulate further theoretical work in SR of small systems
  where noise intensity and thermodynamic stability are tightly coupled.
  Another consequence of the noise intensity vs thermodynamic stability
  coupling is the strong variability exhibited by single-molecule SR
  experiments: the measured signal-to-noise ratio versus any control parameter (in our case, oscillation frequency) will tend to
  show large variations from molecule to molecule.  This was
  apparent in the results for hairpin H1 shown in Fig. 3 and has been
  observed in the rest of molecules (see, for instance the results shown
  in Fig. 7 for SH8). Such variability is consequence of the aforementioned weak
  stability of biomolecular bonds, and various sources of experimental
  errors (e.g. instrumental drift, misalignment attachment, inaccurate
  discrimination of the co-existence force, etc..). It has no counterpart in other
  SR studies of non-linear macroscopic devices or single
  degree-of-freedom systems (such as single colloidal particle in optical traps or
  macroscopic systems in solid state physics or electronic devices).  

\section{Future perspectives}

The results of our work suggest that we could extract the kinetic rates of 
molecular hoppers by measuring the resonance 
frequency in oscillating experiments. Is this approach 
useful? There are several widely accepted and commonly used 
single-molecule methods that can extract the 
kinetic parameters of molecular hoppers just by analyzing the hopping 
traces without bothering about carrying out 
oscillating measurements. It  
is then clear that single-molecule SR is not worth  
pursuing if other simpler methods are available. Yet SR might be of  
interest for investigating fast molecular transitions where current  
methods might fail. In Section VII, we investigated SR in an 8bp short 
DNA hairpin (SH8) at conditions (low trap stiffness) where hopping rates 
are hard to be measured from standard methods (e.g. the Bell-Evans 
model). The faster hopping rate and the smaller jumps in extension (due to 
both the shorter length of SH8 and the 
decreased trap  stiffness) 
contribute to make the hopping rate measurements 
difficult. Note that we have been able 
to extract the value of the hopping rate  
either by measuring the power spectrum (Fig. 6b) or by implementing a 
hidden-Markov model.  
Interestingly, whereas applying 
standard methods to extract kinetic rates become steadily difficult as 
the  hopping   
signal becomes more noisy, the quality of the resonant peak in the 
SNR remains acceptable (Fig. 6d). This suggests that in experimental 
conditions  where hopping signals become nearly undetectable,  
SR may find a fertile ground for useful applications.  

\begin{figure}
\begin{center}
\includegraphics[width=8cm]{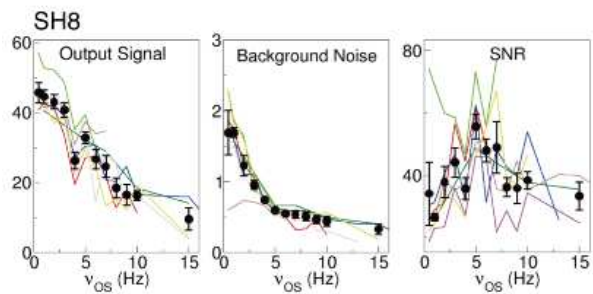}
\end{center}
\caption{ {\bf Molecular variability of the measured responses (SH8).} 
Results for the OS, BN and SNR for 7 different SH8 molecules. 
Units: OS (nm$^2$), BN (nm$^2$/Hz), 
SNR (Hz). The error bars represent the standard error over different molecules.
}
\label{fig7}
\end{figure}

Measuring the kinetics of single bonds might be crucial to dissect
  the kinetic pathways of many reactions, from nucleic acid translocases
  indispensable in virtually all tasks of nucleic acids metabolism, to
  molecular folding of proteins and ligand-receptor binding. Moreover,
  the detection of single bond kinetics also provides a direct
  measurement of the affinity (or free energy of formation) of weak
  single bonds (e.g. important for an accurate determination of the
  parameters characterizing the thermodynamics of secondary structure
  formation in nucleic acids \cite{huguet2010}). It is therefore important to explore new
  approaches capable of illuminating into such questions. The experimental
  resolution of formation/dissociation kinetics is currently limited to
  5 base pairs \cite{design,huguet2009}. Overcoming this limit strongly
  relies not only on increasing the hopping signal relative to the noise
  but also in slowing down the (expected too fast) formation/dissociation kinetics of single
  bonds. A direct
  measurement of the formation/dissociation fast kinetics of single
  molecular bonds stretchable along sub-nanometer scales and resistant
  to low (a few piconewton) forces remains an experimental challenge.

In fact, the route to discriminate hopping kinetics in a small number of base pairs 
may be plagued of difficulties. The situation might be even worse if the aim 
is to detect the unraveling kinetics of a single nearest-neighbor base pair 
(NNBP), which is the minimal unit of DNA bonds (double stranded helices 
are stabilized by both hydrogen bonds between complementary bases 
and stacking between NNBP). 
Currently most kinetics measurements are carried out in hopping experiments.  
However there is a complication 
present in hopping experiments due to the low signal-to-noise ratio 
inherent to unraveling a single NNBP together with the disturbances 
caused by the multifrequential noise present in the high frequency 
range where the kinetic rate of formation/dissociation of a single NNBP 
is expected to fall.  
The low signal-to-noise ratio problem can be partially resolved using 
advanced data analysis tools such as Bayesian methods and HMM to unravel 
hopping traces for SH8. However such methods assume a specific form of 
the noise (i.e. decorrelated force fluctuations and Gaussian emission 
signal) and do not account for multifrequential sources of noise 
(due to the aforementioned sources). In this regard, SR might be 
extremely useful to separate the true formation/dissociation kinetics 
of a single NNBP from these other artifacts.

Finally our work focused in the SR phenomenon in DNA
  hairpins wehereas other interesting molecular
  structures are now available for single molecule pulling. From this
point of view, it would be very interesting to carry out SR  
measurements in more complex molecular folders (e.g. exhibiting  
multiple folding pathways, intermediate states or non-cooperative  
transitions) such as RNAs and proteins. 
\\
\\
{\bf Methods.}  
\\ {\bf Synthesis of DNA hairpins.}  The DNA hairpins
with handles are synthesized using the hybridization of three
different oligonucleotides (Fig. 1a). One oligonucleotide contains the
sequence of the ssDNA left handle plus a part of the sequence of the
desired DNA hairpin; the second has the rest of the sequence of the
DNA hairpin and the ssDNA right handle. The right and the left handles
have the same sequence in order to hybridize them with the third
oligonucleotide. The first oligonucleotide has a biotin at its 5' end
and the second oligonucleotide has been modified at its 3' end with a
digoxigenin tail (DIG Oligonucleotide Tailing Kit, 2nd generation,
Roche Applied Science). Once the first and the second oligonucleotides
are hybridized to form the hairpin, the third oligonucleotide is
hybridized to the handles to form the dsDNA
handles. Streptavidin-coated polystyrene microspheres (1.87~$\mu$m;
Spherotech, Livertyville, IL) and protein G microspheres (3.0-3.4~$\mu$m; 
G. Kisker Gbr, Products for Biotechnology) coated with
anti-digoxigenin polyclonal antibodies (Roche Applied Science) were
used for specific attachments to the DNA molecular constructions
described above.  Attachment to the anti-digoxigenin microspheres was
achieved first by incubating the beads with the tether DNA. The second
attachment was achieved in the fluidics chamber and was accomplished
by bringing a trapped anti-digoxigenin and streptavidin microspheres
close to each other. The sequences of the short hairpins are: SH10
(5'-GCGGCGCCAGTTTTTTTTCTGGCGCCGC-3'), SH8
(5'-GGCGCCAGTTTTTTTTCTGGCGCC-3').
\\
{\bf Experimental setup.}
The experiments have been carried out using a high stability newly 
designed miniaturized dual-beam optical tweezers apparatus \cite{huguet2010}. 
It consists of two counter-propagating laser beams of 845~nm wavelength 
that form a single optical trap where particles can be trapped by gradient 
forces. The DNA hairpin is tethered between two beads (Fig.~1a). One bead 
is immobilized at the tip of a micropipette that is glued to the 
fluidics chamber; the optical trap captures the other bead. The light 
deflected by the bead is collected by two photodetectors located at 
opposite sides of the chamber. They directly measure the total 
change in light momentum which is equal to the net force acting on the 
bead. Piezo actuators bend the optical fibers and allow the user to move 
the optical trap. The force is made to oscillate using a force feedback 
system that operates at 4~kHz minimizing instrumental drift effects as 
compared to protocols without feedback. Force feedback does not introduce  
artifacts in our measurements unless $\nu_{\rm os}$ is too high (typically 
larger than 50~Hz) or $A$ is too small (less than 0.1~pN). 

The folding-unfolding experiments described in this report were performed 
at room temperature (24$^\circ$C)  in a buffer containing 10mM 
Tris-HCl pH7.5, 1mM EDTA, 1M NaCl, 0.01\% Sodium Azide.
\\
\\
\noindent{\bf Acknowledgements.}  K. H. is supported by 
Grant-in-Aid for Scientific Research from the MEXT (No. 23107703). 
M. M. is supported by HFSP RGP0003/2007-C and 
EU (BioNanoSwitch). J. M. H., N. F. and F. R. are supported by grants
FIS2007-3454, Icrea Academia 2008 and HFSP (RGP0055-2008).


\begin{thebibliography}{99}



\bibitem{benzi1}
R. Benzi, A. Sutera  and  A. Vulpiani, 
{\it The mechanism of stochastic resonance}, 
J. Phys. A 
{\bf 14}, L453 (1981). 

\bibitem{benzi2}
R. Benzi, G. Parisi, A. Sutera and  A. Vulpiani,  
{\it Stochastic resonance in climatic changes}, 
Tellus 
{\bf 34}, 10 (1982). 



\bibitem{simon} 
A. Simon  and A. Libchaber,  
{\it Escape and synchronization of a Brownian particle}, 
Phys. Rev. Lett. {\bf 68}, 3375 (1992). 


\bibitem{schmitt} C. Schmitt, B.  Dybiec, P.  H\"anggi and C. Bechinger, 
{\it Stochastic resonance vs. resonant activation},  
Europhys. Lett.   {\bf 74}, 937 (2006). 

\bibitem{ciliberto}
P. Jop, A.  Petrosyan and S. Ciliberto,  
{\it Work and dissipation fluctuations 
near the stochastic resonance of a colloidal particle}, 
Europhys. Lett. {\bf 81}, 50005 (2008). 

\bibitem{bio} 
D. F. Russell, L. A.  Wilkens  and F. Moss,  
{\it  Use of behavioural 
stochastic resonance by paddle fish for feeding}, 
Nature {\bf 402}, 
291-293 (1999). 

\bibitem{McDonAb09} M. McDonnell and D. Abbott, 
{\it What is stochastic resonance? Definisions, Misconceptions, debates, 
and its relevance to biology},
 PLoS Comp. Bio. {\bf 5}, e1000348 (2009).

\bibitem{reviewer2} D. Petracchi, M. Pellegrini, M. Pellegrino, M. Barbi and 
F. Moss, 
{\it Periodic forcing of a K$^+$ channel at various temperature}, 
Biophys. J. {\bf 66}, 1844 (1994). 

\bibitem{quantum1} 
M. Grifoni and P. H\"anggi,  
{\it Coherent and incoherent quantum stochastic
resonance}, 
Phys. Rev. Lett. {\bf 76}, 11 (1996). 


\bibitem{quantum2} 
D. Witthaut, F. Trimborn and S. Wimberger,  
{\it Dissipation-induced coherence 
and stochastic resonance of an open two-mode Bose-Einstein condensate}, 
Phys. Rev. A {\bf 79}, 033621 (2009). 

\bibitem{hanggi} 
L. Gammaitoni, P.  H\"anggi, P. Jung and F. Marchesoni, 
{\it Stochastic resonance}, 
Rev. Mod. Phys. {\bf 70}, 223 (1998). 

\bibitem{wellens} 
T. Wellens, V. Shatokhin and A. Buchleitner, 
{\it  Stochastic resonance}, 
Rep. Prog. Phys. {\bf 67}, 45 (2004). 


\bibitem{Nuria}
N. Forns, S. de Lorenzo, M.  Manosas, K. Hayashi, J. M.  Huguet and F. Ritort,  
{\it Improving signal-to-noise resolution in single molecule experiments 
using molecular constructs with short handles}, 
Biophys. J. {\bf 100}, 1765 (2011). 

\bibitem{footnote1}
For 29bp molecular handles ($\simeq$10~nm of contour length at each 
flanking side), the molecular extension is short ($\simeq 20$nm) and 
the beads are 
very close to each other. However the high stiffness of the handle under 
tension prevents the beads from clashing \cite{Nuria}.

\bibitem{exp2}
J-D. Wen,  M.  Manosas, P. T. X. Li, S.  B. Smith, C.  Bustamante, 
F. Ritort and I. Tinoco, 
{\it  Force unfolding kinetics of RNA using optical tweezers. 
I. Effects of experimental variables on measured results},  
Biophys. J. {\bf 92}, 2996 (2007). 

\bibitem{exp3}
M. Manosas, J.-D. Wen, P.T.X. Li, S.B. Smith, C. Bustamante, I. Tinoco 
and F. Ritort, 
{\it Force unfolding kinetics of RNA using 
optical tweezers. II. Modeling experiments}, 
Biophys. J. {\bf 92}, 
3010 (2007). 

\bibitem{NamWie89} B. McNamara and K. Wiesenfeld, 
{\it Theory of stochastic resonance}, 
 Phys. Rev. A {\bf 39}, 4854 (1989).

\bibitem{Stocks} 
N. G. Stocks, 
{\it A theoretical study of the non-linear response of a periodically 
driven bistable system}, 
Nuovo Cimento D {\bf 17}, 925 (1995).



\bibitem{pre}
A. L. Pankratov,  
{\it  Suppression of noise in nonlinear systems subjected to 
strong periodic driving}, 
Phys. Rev. E {\bf 65}, 022101 (2002). 


\bibitem{snr1}
V. Berdichevsky  and M. Gitterman,  
{\it  Stochastic resonance in a bistable 
piecewise potential: analytical solution},  
J. Phys. A: Math. Gen. 
{\bf 29}, L447 (1996). 

\bibitem{snr2}
M. C. Mahato and A. M. Jayannavar, 
{\it  Some stochastic phenomena in a driven 
double-well system}, 
Physica A {\bf 248}, 138 (1998).

\bibitem{casado} 
J. Casado-Pascual, J. Gomez-Ordo\~nez, M.  Morillo and 
P.   H\"anggi,  
{\it  Two-state theory of nonlinear stochastic resonance}, 
Phys. Rev. Lett. {\bf 91}, 210601 (2003).

\bibitem{p1}
L. Gammaitoni, F. Marchesoni and S. Santucci, 
{\it  Stochastic resonance as a bona fide resonance}, 
Phys. Rev. Lett. {\bf 74}, 1052 (1995). 

\bibitem{p1-2}
T. Zhou, F. Moss  and P. Jung, 
{\it Escape-time distributions of a periodically modulated bistable system with 
noise}, 
 Phys. Rev. A   {\bf 42}, 3161 (1990). 

\bibitem{work}
S. Saika, R.  Ratnadeep and A. M. Jayannavar,
{\it  Work fluctuations and stochastic resonance}, 
 Physics Letters A {\bf 369}, 367 (2007). 

\bibitem{Mossa09} A. Mossa, S. De Lorenzo, J. M. Huguet and F. Ritort, 
{\it Measurement of work in single molecule experiments}, 
J. Chem. Phys. {\bf 130}, 234116 (2009).

\bibitem{jung} P. Jung and F. Marchesoni, {\it Energetics of stochastic resonance}, 
CHAOS {\bf 21}, 047516 (2011).














\bibitem{design} 
M. T. Woodside,   W. M. Behnke-Parks, K. Larizadeh, K. Travers, D.  
Herschlag and S. M. Block, 
{\it Nanomechanical measurements of the 
sequence-dependent folding landscapes of single nucleic acid hairpins}, 
Proc. Natl. Acad. Sci.  {\bf 103}, 6190 (2006).  

\bibitem{Mossa09b} A. Mossa, M. Manosas, N. Forns, J. M. Huguet and 
F. Ritort, 
{\it Dynamic force spectroscopy of DNA hairpins (I): Irreversibility and 
dissipation}, J. Stat. Mech., P02060 (2009).

\bibitem{hmm}
F. E. M\"ullner, S. Syed, P. R. Selvin and F. J. Sigworth, 
{\it Improved hidden Markov models for molecular motors, Part 1: Basic theory}, 
Biophys. J. {\bf 99}, 3684 (2010). 







\bibitem{huguet2009} J. M. Huguet, N. Forns and F. Ritort, {\it Statistical properties of metastable intermediates in DNA
unzipping}, Physical Review Letters, {\bf 103} 248106 (2009).


\bibitem{huguet2010} J. M. Huguet, C. V. Bizarro, N. Forns, S. B. Smith, 
C. Bustamante and F. Ritort, 
{\it Single-molecule derivation of salt dependent base-pair free energies in 
DNA}, 
Proc. Natl. Acad. Sci. {\bf 107}, 15431-15436 (2010). 

\end{thebibliography}
\end{document}